# Quenched Results on $B$ Mesons with NRQCD

Presented by A. Ali Khan[a][*]

[a]Dept. of Physics and Astronomy, University of Glasgow, Glasgow G12 8QQ

Preliminary results on spectrum and decay constant of $B$ mesons from a quenched simulation at $\beta = 6.0$ on an $16^3 \times 48$ lattice are discussed. The heavy quark has been implemented using both an NRQCD and the static formulation, the light one with a clover improved Wilson formulation. Both the NRQCD Hamiltonian and the heavy-light current consistently include terms of $O(1/M_Q)$, the inverse heavy quark mass.

## 1. INTRODUCTION

Calculations of the spectrum of $B$ mesons and the decay constant $f_B$ are of topical interest in lattice gauge theory. Spectrum calculations allow to check the viability of the lattice methods by comparison to experimental results.

In $B$ systems the mass of the heavy quark is $> 1$ in lattice units. If a relativistic action like Wilson or clover is used naively for both the heavy and the light quark one has to simulate at smaller quark masses and then extrapolate the results to the $b$ mass. The static approximation on the other hand, with an infinitely massive heavy quark, is expected to yield too a high value for $f_B$ compared to a $B$ meson with realistic quarks and cannot reproduce heavy mass dependent splittings correctly. In NRQCD the heavy quark is treated in the first approximation as nonrelativistic, and corrections due to relativity are introduced systematically order by order. This methods allows us in principle to study $B$ systems directly, without the need of any extrapolation. For heavy-light systems the parameter with respect to which the action is expanded is given by the inverse heavy quark mass $1/M_Q$.

## 2. SIMULATION

The nonrelativistic Hamiltonian $H$ used here to describe the $b$ quark and the pseudoscalar heavy-light current are corrected through order $1/M_Q^0$, where $M_Q^0$ is the bare heavy quark mass, at tree level. So $H$ looks as follows:

$$H = Q^\dagger \left( D_t + \frac{D^2}{2M_Q^0} \right) Q + Q^\dagger \frac{\sigma B}{2M_Q^0} Q, \quad (1)$$

where Q is the two component heavy quark spinor. The Hamiltonian is tadpole improved, i. e. the gauge fields are divided by the fourth root of the average plaquette:

$$U_\mu \to U_\mu/u_0, \quad u_0^4 = \langle \tfrac{1}{3} Tr U_{\text{Plaq.}} \rangle. \quad (2)$$

Also the heavy-light current is corrected through $O(1/M_Q)$. This is done by considering the Foldy-Wouthuysen transform which relates the small components of the heavy quark 4-spinor $q_h$ to the non-relativistic spinor $Q$ to the desired order in $M_Q$.

$$q_h \simeq (1 - iS^{(0)})Q, \quad S^{(0)} = -\frac{i}{2M_Q^0} \gamma \cdot D \quad (3)$$

We extract $f_B$ from a matrix element of the temporal component of the axial vector current between a pseudoscalar state i.e. the $B$ meson, and the vacuum. This matrix element in NRQCD is related as follows to its counterpart in relativistic QCD:

$$\langle \bar{q} \gamma_5 \gamma_4 q_h | PS \rangle_{QCD} = \langle q^\dagger Q | PS \rangle_{NRQCD} +$$
$$\frac{1}{2M_Q^0} \langle q^\dagger (-i\sigma \cdot D) Q | PS \rangle_{NRQCD} \quad (4)$$

Here, a bare heavy quark mass of $M_Q^0 = 1.71$ in lattice units is used.

This computation was performed on a set of 36 quenched configuration at $\beta = 6.0$ on a $16^3 \times 48$

lattice, fixed to Coulomb gauge. For light quarks we use clover improved Wilson fermions with the clover coefficient $c = 1$, i.e. lattice spacing errors removed through $O(a)$ at tree level, at $\kappa$ values of 0.1432 and 0.1440. The critical and strange $\kappa$ values for these propagators are $\kappa_c = 0.14551(3)$ and $\kappa_s = 0.1437(1)$, the scale determined from $M_\rho$ is $a^{-1} = 2.05(6)$ GeV [1]. Here we would like to point out that for quenched configurations the scale determination does vary heavily depending on the physical quantities considered, as expected. The scale obtained from $\Upsilon$ spectroscopy at $\beta = 6.0$ is 2.4(1) GeV [3].

We studied s wave correlation functions in pseudoscalar and axial vector channels, both local and smeared at source and sink. The smearing functions used in our run and at SCRI [2] are hydrogen-like wavefunctions (ground and 1st excited state for the NRQCD propagators and ground state for the static ones).

## 3. ANALYSIS AND RESULTS

### 3.1. B and $B_s$ masses

In NRQCD the meson mass $m$ can be calculated by adding the nonrelativistic binding energy which is extracted from the heavy-light correlation functions and an energy shift:

$$m = \Delta + E_{\text{NRQCD}} \qquad (5)$$

The energy shift $\Delta$ contains the renormalized quark masses and the zero point of the energy. Both can be calculated perturbatively [3]. The $E_{\text{NRQCD}}$'s shown in table 1 have been extracted from fits of the ground and excited state smeared-local correlators to 2 exponentials. An extrapolation of the results to $\kappa_c$ and $\kappa_s$ yields 0.437(5) and 0.491(6) respectively. Converting this into physical units we get a $B - B_s$ mass splitting of 0.11(3) GeV, compared to the experimental value of 0.096(6) GeV. In Heavy Quark Effective Theory (HQET) this splitting is expected to be independent of $M_Q$. For a further discussion of mass splittings in the context of HQET see [2]. For the B and $B_s$ masses themselves we get 4.5(4) and 4.6(4) GeV respectively. This difference with experiment may be accounted for by the fact that we chose $M_Q^0$ in lattice units to give the right $\Upsilon$ mass assuming that the lattice spacing was 2.4 GeV, the appropriate value for $b\bar{b}$ systems. The dependence of $a^{-1}$ on the scale in the quenched approximation means that we have to retune $M_Q^0$ for B mesons, where we believe that the light scale is more appropriate for most physical phenomena.

Table 1
Results in lattice units

| $\kappa$ | 0.1432 | 0.1440 |
|---|---|---|
| $E_{\text{NRQCD}}$ | 0.506(6) | 0.482(5) |
| $\Delta E$ ($B^* - B$) | 0.021(1) | 0.021(1) |
| $Z_L$(static) | 0.238(7) | 0.225(2) |
| $Z_L$(NRQCD) | 0.120(5) | 0.113(5) |

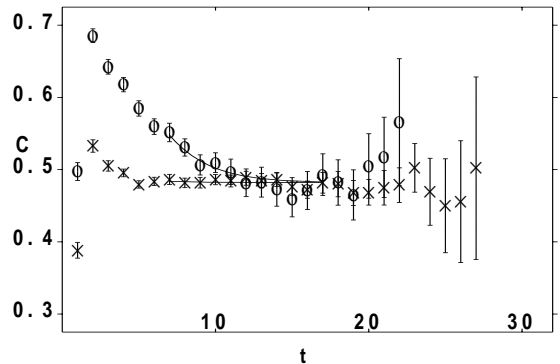

Figure 1. Smeared-local correlation functions (crosses: ground state, circles: first excited state) with 2-exponential fit from t = 7 to t = 17. $\kappa = 0.1440$.

### 3.2. $B^* - B$ splitting

To extract the energy splitting $\Delta E$ between the $^3S_1(B^*)$ and the $^1S_0(B)$ state we fit the bootstrapped ratio of the smeared-local correlators to a single exponential:

$$\frac{C_{SL}(^3S_1)}{C_{SL}(^1S_0)} \propto A \exp(-\Delta E t). \qquad (6)$$

We obtain $\Delta E = 0.021(1)$ for both $\kappa$ values. From HQET one expects this splitting to be proportional to $1/M_Q$, so it is not only directly dependent on $a^{-1}$ but also through the value of the heavy mass in lattice units. Having chosen too small a value for $M_Q^0$ for our simulations we rescale $\Delta E$ when converting it into physical units, which yields a splitting of 37(6) MeV, compared to the experimental value 46(1) MeV. If one assumes that $\Delta E$ is a short distance quantity, one might argue that a higher scale is more appropriate here. Using $a^{-1} = 2.4(1)$ GeV one would obtain 50(4) MeV in physical units.

### 3.3. $f_B$

The continuum value of $f_B$ is related to the corresponding lattice quantity $Z_L$ through

$$f_B \sqrt{M_B} = \sqrt{2} Z_L Z_A a^{-3/2}. \qquad (7)$$

For sufficiently large times the ratio of the correlation functions

$$C_{SL}/C_{SS} \sim Z_L/Z_S, \qquad (8)$$

So the first step to determine $Z_L$ is to fit the bootstrapped ratio of the two correlation functions, starting from the time slice where its plateau begins, to a constant. $Z_S$ is obtained from a single exponential fit to the ground state smeared-smeared correlator:

$$C_{SS} \sim Z_S^2 \exp(-Et) \qquad (9)$$

The results for $Z_L$ in NQRCD and static cases are given in table 1. Extrapolating to the critical and strange $\kappa$ values gives 0.103(4) and 0.115(5) for NRQCD and 0.201(6) and 0.230(7) in the static case. In figure 2 we show the ratio of the small components contribution to the first term in equation 4, which is about 20%. There is no value for $Z_A$ in NRQCD available yet, whereas for the static theory there is a perturbative result of 0.78. If we assume that the light scale would be most appropriate we obtain in the static approximation $f_B = 0.28(3)$ GeV and $f_{B_s} = 0.32(3)$ GeV, which agrees with the UKQCD value [4] within errors, as expected. With $a^{-1} = 2.4(1)$ GeV we get the substantially higher values of $f_B = 0.36(3)$ GeV and $f_{B_s} = 0.41(4)$ GeV.

### 4. CONCLUSIONS

The good signal for our correlation functions and well chosen s state smearing functions enable us to extract NRQCD energies with a good accuracy. The conversion of results to physical units suffers from a large uncertainty in the scale; in our case the main source of this uncertainty is quenching. Only given an accurate value for $a^{-1}$ we may be able to calculate energies and $f_B$ accurately. We find that there are large $1/M_Q$ corrections for $Z_L$ when going from a static to an NRQCD heavy quark where Hamiltonian and axial vector current are corrected through $O(1/M_Q)$. Once $Z_A$ has been calculated in NRQCD it will be seen how this translates into the $1/M_Q$ dependence of $f_B$.

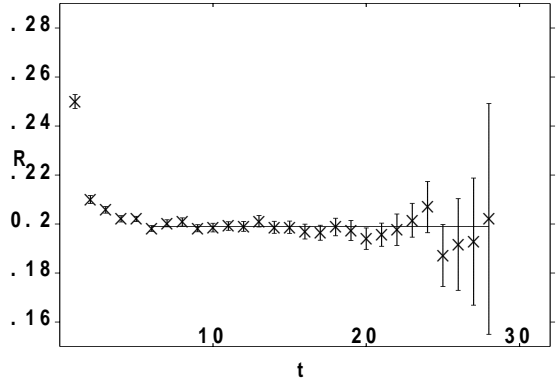

Figure 2. Ratio **R** of the small components correction to the pseudoscalar correlation function. The solid line shows a fit of the ratio to a constant from t = 6 to t = 28.

We would like to thank the UKQCD collaboration for allowing us to make use of their configurations and light propagators. In particular we have been helped by H. Shanahan and H. Newton.

This work was supported by SHEFC, PPARC and the EU under contract CHRX – CT92 – 0051.